# A Comparative Study of Protein Unfolding in Aqueous Urea and DMSO Solutions: Surface Polarity, Solvent Specificity and Sequence of Secondary Structure Melting

**Susmita Roy and Biman Bagchi***

Solid State and Structural Chemistry Unit, Indian Institute of Science, Bangalore 560012, India.

*Email: bbagchi@sscu.iisc.ernet.in

***ABSTRACT***

**Elucidation of possible pathways between folded (native) and unfolded states of a protein is a challenging task, as the intermediates are often hard to detect. Here we alter the solvent environment in a controlled manner by choosing *two different co-solvents* of water, urea and dimethyl sulphoxide (DMSO), and study unfolding of *four different proteins* to understand the respective sequence of melting by computer simulation methods. We indeed find interesting differences in the sequence of melting of alpha-helices and beta-sheets in these two solvents. For example, at 8M urea solution, beta-sheet parts of a protein is found to unfold preferentially, followed by the unfolding of alpha helices. In contrast, 8M DMSO solution unfolds alpha helices first, followed by the separation of beta-sheets for majority of proteins. Sequence of unfolding events in four different $\alpha/\beta$ proteins and also in chicken villin head piece (HP-36) both in urea and DMSO solution demonstrate that the unfolding pathways are determined jointly by relative exposure of polar and non-polar residues of a protein and the mode of molecular action of a solvent on that protein.**

## I. Introduction

Mechanism of folding and unfolding of different proteins under different environments is a highly complex physicochemical process. A native state of protein is stabilized and held together in aqueous environments by variety of forces, such as hydrogen bonding and hydrophobic attraction among amino acid residues. Nature of these forces is different. While hydrogen bonding is pair specific, hydrophobic force is more collective, gets enhanced in an aggregate. These forces together determine the pathways of folding and unfolding in a given environment. Thus, the pathway can change as the environment changes and this is an important aspect of protein folding problem.

Folding funnel paradigm with its associated energy landscape view has been applied to find a general semi-quantitative approach to rationalize a vast amount of information available now. [1-20] A major prediction of folding funnel paradigm is the existence of multiple pathways during transition from the folded to the unfolded state and vice versa, and in the process a protein is predicted to evolve through different intermediate states. This prediction has been a subject of intense debate with examples both in favor and in opposition. [1-25] The issue is still far from settled. It is indeed a daunting task to map out the full folding pathway at an atomic level. One possible approach is to use different chemical denaturants in the hope of finding different pathways of unfolding. [4-7] Early work of Daggett and co-workers demonstrated for chymotrypsin inhibitor-2 that in 8M urea solution the β-structure melts first, followed by α-helix. [26] In similar recent studies by Rocco and co-workers compared the unfolding pathways of protein L in different denaturation modes, such as temperature, urea and guanidinium chloride.

[27, 28] They reported that in 10M urea β-sheet is destabilized first, whereas in temperature and 5M GdmCl, it is the α-helix. Englander carried out a series of unfolding studies and concluded that unfolding follows a definite predetermined pathway. [29-31] Fayer and co-workers found the system to settle in a molten globule state. [32] For urea mediated denaturation, two mechanisms have been proposed: (i) An indirect mechanism where urea drags out the hydrogen-bound water molecules by dehydrating the protein surface which indirectly facilitates the unfolding event [33] and, (ii) a direct mechanism where urea preferentially bind with protein molecules through a strong dispersion interaction competing with water. [34, 35]

The fact that most of recent theoretical and molecular dynamics simulations employ urea and guanidine chloride as denaturating solvents has somewhat limited our data pool; [26-28, 33-39] examples of the use of other co-solvents induced protein unfolding are sparse. Among those co-solvents dimethyl sulphoxide (DMSO) is unique. [40-43] It is popularly used as a stock solution in drug discovery processes and efficiently plays roles as a stabilizer, an activator, an inhibitor, a cryoprotector and also as a capable denaturant. While at low concentrations of DMSO ($X_{DMSO}$<0.05) majority of proteins are found to be conformationally unaffected, they undergo a number of structural changes as DMSO concentration increases. We recently carried out studies aimed at understanding quite different molecular interaction that DMSO provides between a protein and its chemical environment. [40] Along with our early studies, many other research groups also found the sensitivity of secondary structure of protein upon aqueous DMSO treatment. [40-44] Very recently, from Raman optical activity measurements, Blanch and coworkers found DMSO to selectively act on alpha helices for a number of proteins at higher DMSO mole fraction whereas beta sheets remained unaffected regardless of solvent concentration. [45] All these works greatly motivate us to examine the capability of DMSO as a helix breaker and the origin of its selectivity at a molecular level.

In the present work we study and compare the unfolding of four proteins using several different denaturating conditions, specifically in various concentrations of aqueous urea and aqueous dimethyl sulphoxide (DMSO) solutions. This allows us to quantify the sensitivity of different secondary structural segments of a particular protein towards addition of such co-solvents. In addition, we selectively choose those proteins where the two principal secondary structural contents, i.e., alpha helix and beta sheet parts present in their native conformation. Unfolding of four such well characterized α/β proteins namely, (1) single immunoglobulin binding domain Protein G (PDB ID: 1GB1) from group G Streptococcus, (2) Chymotrypsin Inhibitor 2 (PDB ID: 2CI2), (3) IgG binding domain of Protein L (PDB ID: 2PTL) and (4) human erythrocytic ubiquitin (PDB ID: 1UBQ) were studied both in aqueous urea and DMSO solutions by molecular dynamic simulation method. We also studied chicken villin head piece HP-36 earlier but that has only helices. [43] The details of molecular dynamic simulation technique are described in the System Setup and Simulation Details section.

## II. Solvent Sensitivity of Secondary Structures in 8M Urea and 8M DMSO
### A. Differential Stability of Alpha Helices and Beta Sheets

To characterize the effects of urea and DMSO on those selected proteins we individually track the time progression of unfolding by following their conformational changes in the particular secondary structural segments. We observe that quite generally urea solution preferentially initiate unfolding of its beta-sheet part of such α/β proteins which is followed by unfolding of the alpha helical content. This is also in accordance with other simulation results. [26-28] In contrast, the 8M DMSO solution initially unfolds alpha helix and then beta sheet, with lower priority. DMSO induced such selective melting has also been observed by recent Raman optical activity measurements. [45] Figure 1 shows diverse effects of urea and DMSO on a single protein, GB1 (Protein G). In Figures 1(a) and (b), we display the time evolution of the fraction native contact ($\eta$) exclusion/inclusion dynamics. At 8M urea solution, while a considerable amount of correct native contact formation of beta sheet segment is diminished within 10ns of timescale (to $\eta = 0.4$), that of alpha-helix remains larger than 0.9 at the same timescale, which eventually starts to disappear long time after (20ns) the initial unfolding event. At 8M DMSO solution, in contrast, alpha helical native contacts break faster than that of beta sheet. Time progression up to 50ns shows that the fraction of native contact ($\eta$) decreases only up to 0.6 in case of beta sheet segment of protein G. The contact map analysis in the transition state of unfolding clearly provides the evidence of disappearance of alpha helical and beta sheet segments in 8M DMSO and 8M urea, respectively (see Figures 1(c) and (d)). The emergence of two different intermediates (as shown in Figure 1(e)) during the unfolding of Protein G in 8M urea and 8M DMSO provide the most important evidence for the existence of two different denaturation pathways of unfolding. Conformational fluctuation of protein G in terms of root mean square fluctuation of alpha helix and beta sheet residues also show distinct behaviour in 8M urea and 8M DMSO (see Figure S1). While destabilization phenomenon of secondary structures in terms of correct native contact arrangement dramatically differs for a specific protein in these two solvents, late stage enhancement of radius of gyration ($R_g$) remains a common event in the unfolding of protein G in any of these two solvents (see Figure 1(f)). This signifies the final splitting of tertiary structure which causes elongation of the whole polypeptide chain.

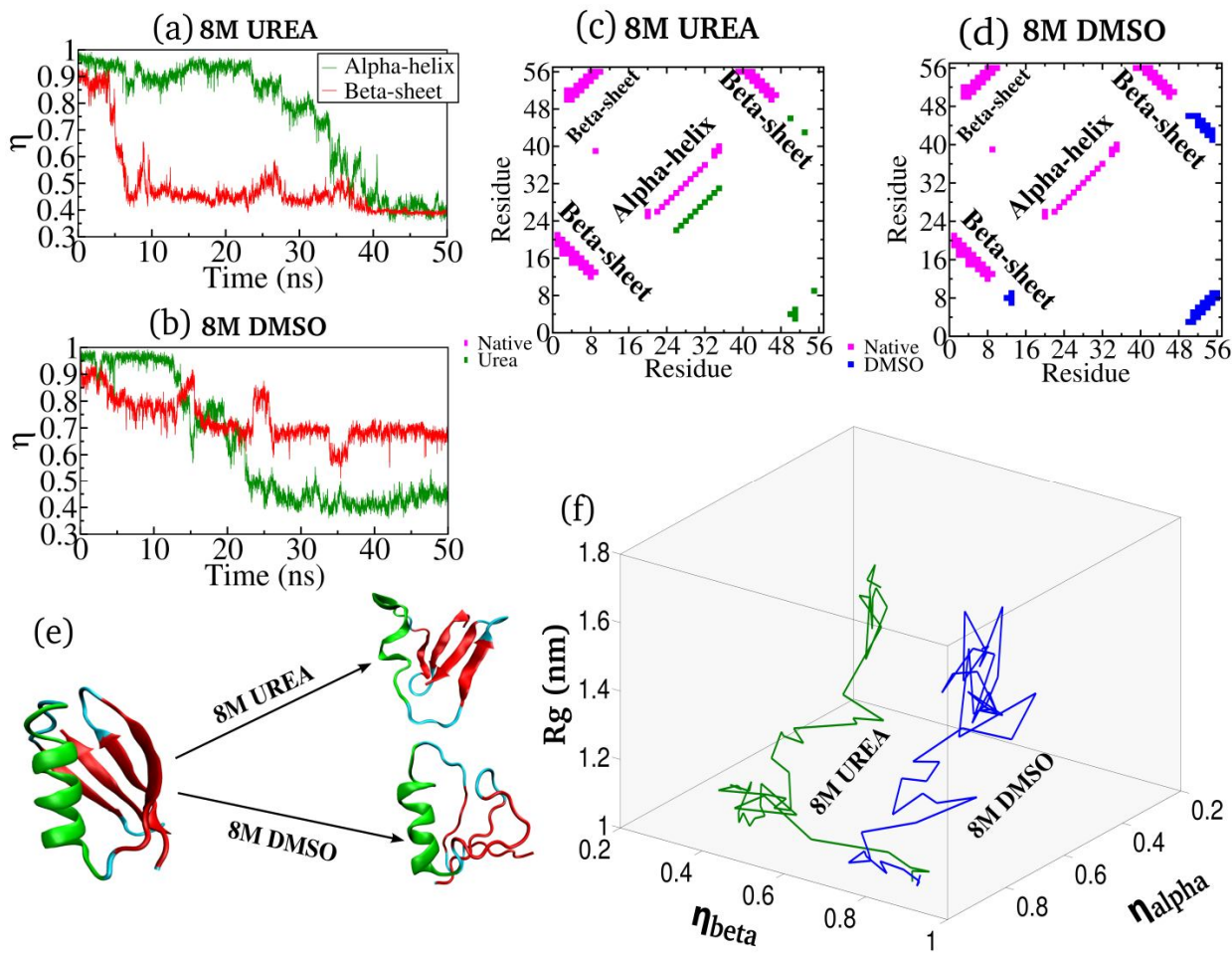

**Figure 1.** Fraction of native contact ($\eta$) dynamics of protein G, in (a) 8M urea (b) 8M DMSO solutions. Contact map of protein G at intermediate simulation time frame (20ns) in (c) 8M urea, (d) 8M DMSO. (e) Snapshot of two different intermediates obtained from urea and DMSO induced unfolding trajectories. (f) Protein's conformational degrees of freedom along fraction of beta sheet native contact (X-axis:$\eta_{beta}$), fraction of alpha helix native contact (Y-axis:$\eta_{alpha}$), and along the radius of gyration (Z-axis: Rg) at their explicit time.

## B. Property Based Free Energy Contours Detect Two Distinct Pathways

To clearly demonstrate different unfolding trajectories, we derive a property based free energy path contour of protein G during unfolding in 8M urea and 8M DMSO solutions (see Figure 2(a) and (b)). This contour plot shows that the transition from native folded state to unfolded structure follows a minimum energy pathway evolving through different intermediates. It also provides the free energy separation between the folding and unfolding minima for both the two solvents. We could identify the ensemble of partially unfolded intermediates that are different in conformation and essentially bear the signature of different pathways in aqueous urea and DMSO solutions. It is worthwhile to note that unfolding by temperature and also by using GdmCl denaturants reveal diversity in the unfolding pathways of the structurally similar proteins.[26-28] Hence one can surmise that the protein unfolding steps under any denaturing condition proceeds by following a specific minimum energy path which appears to be dominant among the multiple pathways of unfolding. That dominant route however might differ depending on the mode of denaturation.

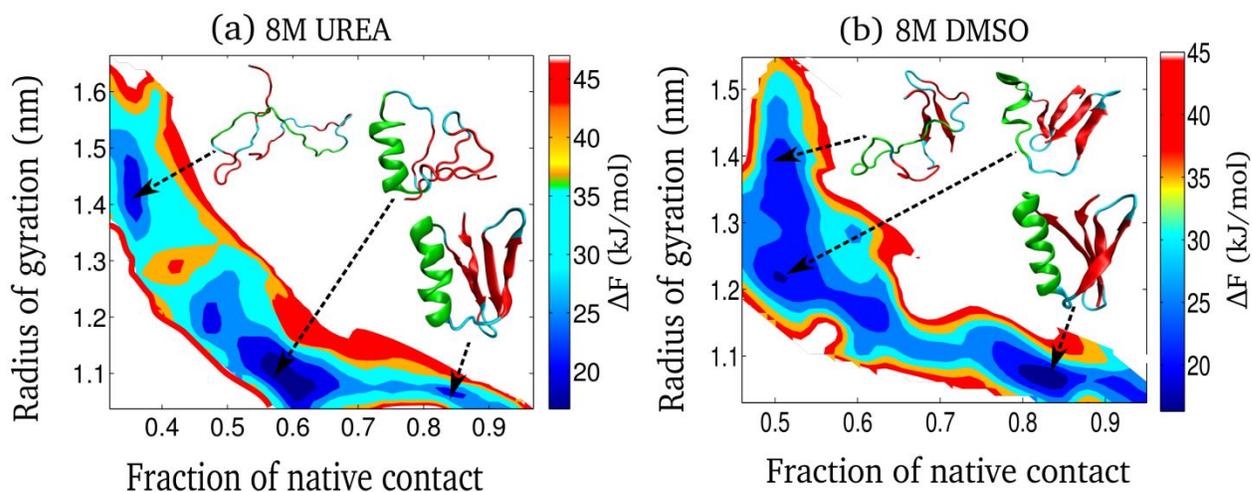

**Figure 2.** Radius of gyration (Rg) and contact order parameter ($\eta$) based free energy landscape along with relatively stable intermediates detected from the folding to unfolding transition of protein G at (a) 8M urea and (b) 8M DMSO solution.

**C. Molecular Mechanism of Urea and DMSO-Induced Unfolding Processes**

To study the microscopic mechanism responsible for the emergence of such diverse unfolding events in urea and DMSO, we track their molecular interaction with proteins. In urea, due to the presence of more hydrophilic ends (C=O and two $NH_2$), there is a high propensity of forming hydrogen bonds with the side-chain residues and the backbone of beta-sheet, leading to the preferential binding with beta sheets than alpha helices (see Figure 3(a)). The molecular mechanism of DMSO induced unfolding process is entirely different. It can be attributed to the preferential solvation of the hydrophobic side chain atoms through the methyl groups of DMSO (see Figure 3(b)), followed by the hydrogen bonding of the oxygen atom of DMSO to the exposed backbone NH groups of protein G (see Figure 3(c)). It is important to note that alpha helix of protein G has a relatively large hydrophobic solvent accessible surface area than that of beta sheet. This accounts for the hydrophobic solvation through the methyl groups of DMSO leading to the fast melting of alpha helices.

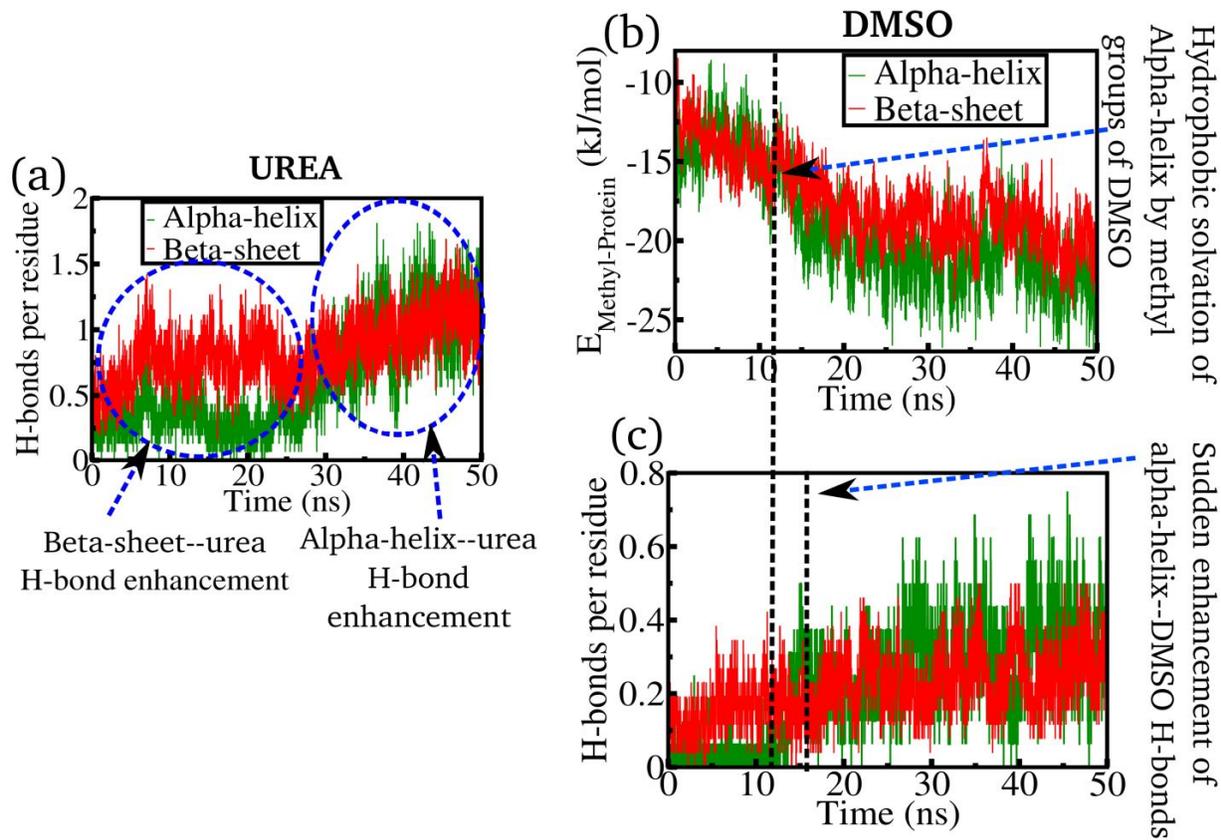

**Figure 3.** (a) Role of protein-urea hydrogen bond interaction in elongation of alpha-helix and beta sheet. (b) Role of the hydrophobic side chain–DMSO interaction and backbone-DMSO interaction underlying the unfolding mechanism.

### D. Solvent Dependence of Unfolding Pathways for Other Alpha-Beta Proteins

We have mentioned earlier that DMSO has attained less attention as a denaturant. To affirm the proposition of microscopic unfolding mechanism induced by DMSO, we selectively study four such α/β proteins including protein G. It also allows us to check whether they share the similar unfolding pathways passing though the alpha helix melted intermediate in 8M DMSO. Combination of radius of gyration and the fraction of native contact dynamics (separately shown in Figure S2(a)-(d)) in terms of their property based free energy path shows that similar to protein G, chymotrypsin inhibitor 2, and protein L also adopt the alpha-helix melted pathway (see Figure 4(a)-(c)). Exception still remains in the crowd. We find maximum unfolding trajectories of ubiquitin in DMSO do not proceed via alpha-helix melted pathway (see Figure 4(d)).

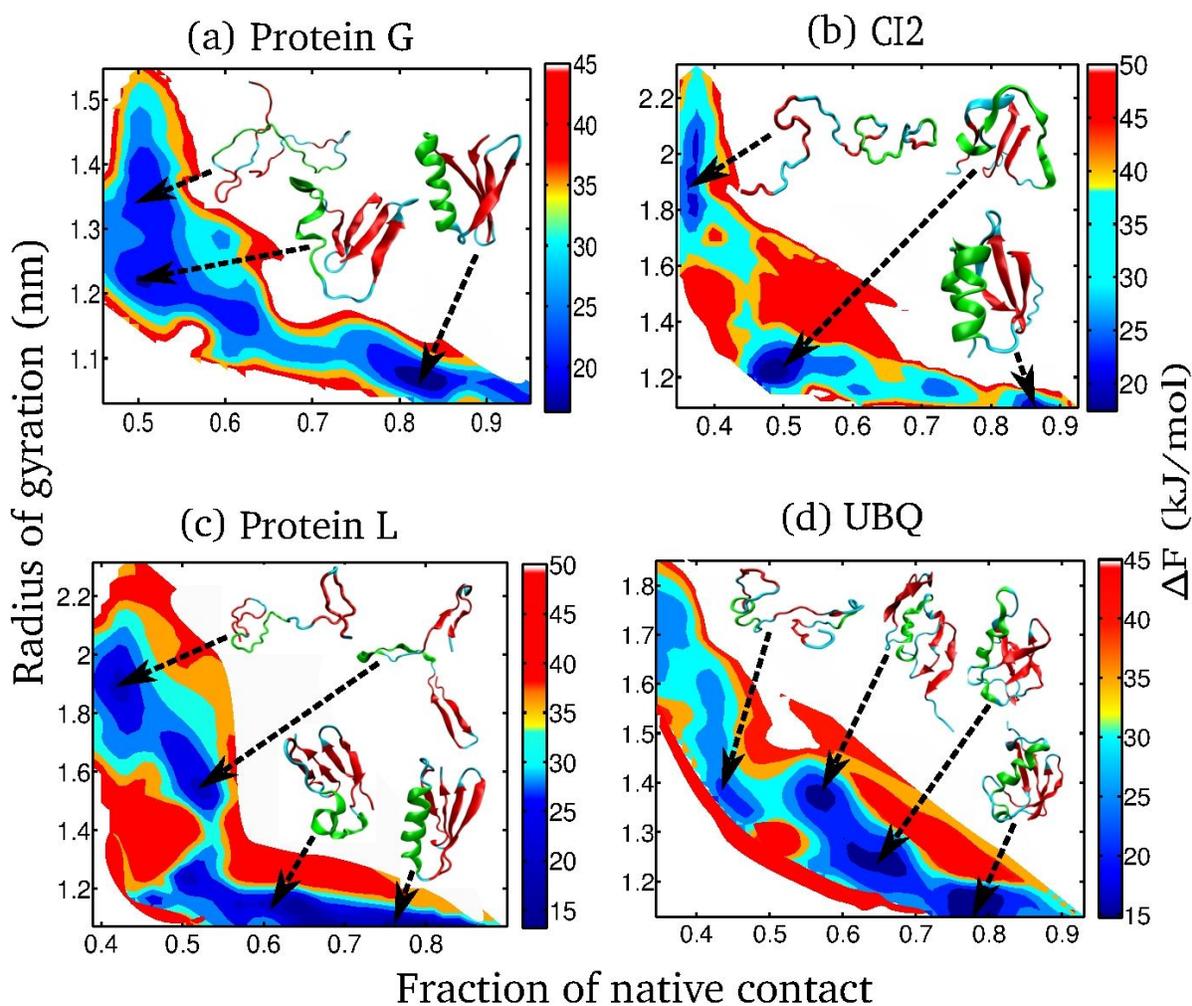

**Figure 4.** Radius of gyration (Rg) and contact order parameter ($\eta$) based free energy landscape along with relatively stable intermediates detected from the folding to unfolding transition of different proteins: (a) Protein G, (b) chymotrypsin inhibitor 2 (CI2), (c) protein L (d) ubiquitin (UBQ), in 8M DMSO solution.

## E. Mode of Action of a Solvent and Polar and Non-Polar Solvent Accessible Surface Area of a Protein together Determine the Dominant Route of Unfolding

Recall that DMSO induced unfolding process is primarily governed by the preferential solvation of the hydrophobic side chain atoms through the methyl groups of DMSO. To estimate the exposure of such side-chain atoms we calculate the relative polar and nonpolar solvent accessible surface area (SASA) (estimates are given in Table 1) for each protein in their crystal structure that can be accessible for solvents we used. We find that all the three proteins, namely protein G, chymotrypsin inhibitor 2, and protein L have relatively higher nonpolar surface exposure of their alpha helical content than the beta sheet which actually leads to the more effective hydrophobic interaction with the methyl groups of DMSO (see Table 1). In addition the residual level analyses reveal that alpha helix enriched with alanine residues also facilitate the alpha-helix melting by preferential hydrophobic solvation. Ubiquitin, in comparison, has relatively less nonpolar surface exposure in its alpha helical segment which might be a dominant factor for choosing a selective unfolding route. This prerequisite information of surface polarity thus can provide immense information about the preferential unfolding pathway of a specific protein in a specific solvent if the molecular mechanism of protein-solvent interaction is already known.

Table 1: Relative polar and nonpolar SASA (solvent accessible surface area) for the alpha helix and beta sheet of four proteins.

|  | Polar SASA | | Non-polar SASA | |
| --- | --- | --- | --- | --- |
|  | Alpha Helix | Beta Sheet | Alpha Helix | Beta Sheet |
| Protein G | 39.30% | 43.80% | 60.70% | 56.20% |
| Chymotrypsin Inhibitor 2 | 31.23% | 40.61% | 68.77% | 59.39% |
| Protein L | 33.50% | 40.75% | 66.50% | 59.25% |
| Ubiquitin | 44.30% | 39.38% | 55.70% | 60.62% |

## III. Conclusion

While folding sequence is sometimes dictated by the local "nativeness" of the unfolded state, there does not seem to exist such a general principle to predict sequence of events unleashed during protein unfolding induced by chemical denaturants. [46] During unfolding a long lived kinetic intermediate can form by long range intra-residual interactions [47-49] and also by apolar-amino acid group – hydrophobic solvent interaction. [46] Such microscopic phenomena effectively dictate the course of unfolding.

Despite the well known facts that both chemical and thermal denaturations unfold a protein, our understanding of the diversity of unfolding mechanism at various environments has still remained at its infancy. Our molecular dynamics simulations aim at elucidating role of different molecular interactions between a protein and its environment that could assist in deciding the choice of a suitable solvent media in many experimental studies in enzymology and protein research. It is, in this spirit, that we propose use of DMSO as a denaturant to melt helices preferentially.

Protein denaturation by aqueous urea gets initiated by the preferential solvation of the polar/charged hydrophilic residues on the protein surface by the polar head groups of urea. This leads to an effective repulsion between the residues on the surface of proteins. As a result, protein swells and its buried hydrophobic residues become exposed. These sequences of events match well with the early explanation reported by Thirumalai and coworkers. [34] In contrast, in aqueous DMSO, protein denaturation occurs by the combined effects of hydrophobic and hydrophilic interactions, *initiated first by the preferential solvation of the nonpolar amino acid residues on the protein surface by the methyl groups of DMSO.* Subsequently, the >S=0 bonds of DMSO that now point outwards from the protein surface, form hydrogen bonds with water. This also gives rise to a swelling of the protein that in turn enhances the availability of backbone to further solvation. Then the >S=0 groups of DMSO pull the backbone outwards through hydrogen bond interaction.

In the present context, for majority of proteins, alpha helices have the larger hydrophobic surface exposure than that of beta-sheets, except ubiquitin. Hence, the majority of proteins choose *preferential hydrophobic solvation* of these helices as the unfolding pathway in DMSO. Our early observations and the current study suggest possible use of DMSO as an excellent helix breaker. On the other hand, majority of proteins having beta sheets with excess hydrophilic surface exposure undergo the conformation changes in aqueous urea by preferentially solvating the polar residues.

However, it is important to note that it may not be true always that urea breaks beta sheets and DMSO melts alpha helices for all the proteins. [26-28, 45] Our analysis differs from the previous ones in this area. An additional merit of the present study is that it demonstrates and explains the origin of such exceptions (as in ubiquitin) , *as a complex interplay between the relative surface exposure of hydrophobic groups in different secondary structures and the polarity and hydrogen bonding ability of co-solvents.*

## IV. System Setup and Simulation Details

As earlier mentioned in the text we selected four well characterized α/β proteins namely, (1) single immunoglobulin binding domain Protein G (PDB ID: 1GB1) [50] from group G Streptococcus, (2) Chymotrypsin Inhibitor 2 (PDB ID: 2CI2), [51] (3) IgG binding domain of Protein L (PDB ID: 2PTL) [52] and (4) human erythrocytic ubiquitin (PDB ID: 1UBQ) [53]. We performed molecular dynamics (MD) simulations of these four proteins in water-urea and water-DMSO binary mixture varying urea and dimethyl sulfoxide (DMSO) compositions by using the GROMACS Package (Version 4.0.5) with the Gromos96 force field (ffG43a1). [54-56]

Numerous evaluations suggest that GROMOS force field can be successfully applied for simulating several biomolecular systems including a number of solvents such as water, urea, chloroform, methanol, DMSO, carbon tetrachloride, etc. Gunsteren and coworkers tested the compatibility of the molecular model for urea with the simple point charge model for liquid water for protein denaturation studies. They have validated molecular dynamics (MD) simulation results to experimental data at 298 K as a function of urea mole fraction. In addition thermodynamic properties, such as density, enthalpy of mixing, free enthalpy of urea hydration, and urea diffusion show well agreement with the experimental values. [57]   Similarly Oostenbrink and coworkers reported that the united atom model of DMSO combined with the GROMOS force field is reliable to produce any physical properties of liquid DMSO, including rotational correlation time, thermal expansion coefficient, isothermal compressibility, specific heat, excess Helmholtz free energy, static dielectric permittivity, shear viscosity, to name a few. All results were in good agreement with experiments. [57]

We performed standard MD simulations of timescale around 50 ns at different composition of urea and DMSO. We essentially practiced the protocol as mentioned by Rocco and co-workers. [27, 28] To accelerate the unfolding process and for better sampling, also in order to avoid the traps in the path we performed a number of MD simulations in water, urea and DMSO at four different temperatures (300K, 350K, 400K, 450K and 480K). We find that temperatures for 300K, 350 K, 400 K, native structures of all the proteins taken were minimally perturbed. They essentially unfolded in water within the range of 450-480K or above. However at 400K temperature, addition of 8M urea or 8M DMSO greatly enhance the rate of unfolding of these proteins. Thus rate of enhancement of unfolding process allowed us to compare several properties of four proteins in urea and DMSO solution within the timescale of 50ns. We monitored several parameters to investigate the different unfolding mechanisms experienced by these proteins under these specific conditions. It is important to note that the residual level secondary structural information for each protein was collected from their pdb source files. [50-53]

In addition to make composition dependent binary mixtures, at first we prepared water-urea and water-DMSO binary mixtures at various concentrations in cubic boxes, with sides 3.0 nm. We used SPC/E model for water molecules. [58] Methyl groups of DMSO were modeled as united atom within ffG43a1 force field. [54-56] After steepest descent energy minimization, each trajectory was propagated in an NVT ensemble and equilibrated for 2 ns. All the simulations in this study were done at 300 K and 1 bar pressure. The temperature was kept constant using the Nose-Hoover thermostat. [59, 60] It was followed by an NPT equilibration for 20 ns using the Parinello–Rahman barostat. [61] After preparing the binary solvents at various concentrations, the selected protein was dissolved to each of them and again followed the same procedure of energy minimization. A total of around 5000-6000 solvent molecules were taken. To further equilibrate the solvent before starting a full molecular dynamics simulation, we hold the protein fixed while allowing the solvent to move around at constant temperature by performing position restrained molecular dynamics for 5ns. This allows the solvent to relax to a state which is natural for the current (native) conformation of the protein. Finally, production runs were performed for each system in an NVT ensemble. All the results were extracted from the 50 ns trajectory. The box size was enlarged to 6-6.5nm to accommodate all the molecules. Periodic boundary conditions were applied and non-bonded force calculations were employed a grid system for neighbor searching. [62] Neighbor list generation was performed after every 10 steps using a cut-off 0.9nm. A cut-off radius of 1.2 nm was used for van der Waal's interaction. To calculate the electrostatic interactions, we used PME with a grid spacing of 0.12 nm and an interpolation order of 4. [63-65]

**Supporting Information**: Figure S1 showing conformational fluctuation of protein G, Figure S2 showing fraction of native contact (η) dynamics, are included in the supporting information. This material is available free of charge via the Internet at http://pubs.acs.org."

**Corresponding Author**

*Email: bbagchi@sscu.iisc.ernet.in

**Notes**

The authors declare no competing financial interests.

**ACKNOWLEDGMENT**

We thank Ms. Rikhia Ghosh for many useful discussions. This work was supported in parts by grants from DST, India. BB acknowledges support from JC Bose fellowship from DST, India.

**REFERENCES**

1. Levinthal, C. Are there pathways for protein folding? *J. Chim. Phys.* **1968**, 65, 44-45

2. Levinthal, C. How to fold graciously. In: "Mossbauer Spectroscopy in Biological Systems." Proceedings of a meeting held at Allerton house, Monticello, Illinois. De-Brunner, P.; Tsibris, J.; Munck, E. (eds.). Urbana, IL University of Illinois Press, **1969**, 22-24

3. Zwanzig, R.; Szabo, A.; Bagchi, B. Levinthal's Paradox. *Proc. Natl. Acad. Sci. U. S. A.* **1992**, 89, 20–22

4. Bryngelson, J. D.; Wolynes, P. G. Intermediates and Barrier Crossing in a Random Energy Model (with Applications to Protein Folding). *J. Phys. Chem.* **1989**, 93, 6902-6915

5. Leopold P. E.; Montal, M.; Onuchic, J. N, Protein Folding Funnels: A Kinetic Approach to the Sequence-Structure Relationship. *Proc Natl Acad Sci U S A.* **1992**, 89, 8721–8725

6. Bryngelson, J. D.; Onuchic, J. N., Socci, N. D.; Wolynes, P. G. Funnels, Pathways, and the Energy Landscape of Protein Folding: A Synthesis. *Proteins Struct. Funct. Genet.* **1995**, 21, 167-195

7. Onuchic, J. N.; Wolynes, P. G. Theory of protein folding. *Curr. Opin. Struct. Biol.* **2004**, 14, 70-75

8. Chan, H.S.; Dill, K.A. Polymer Principles in Protein Structure and Stability *Annu. Rev. Biophys. Biophys. Chem.* **1991**, 20, 447-490

9. Dill, K.; Chan, H. S. From Levinthal to Pathways to Funnels. *Nat. Struct. Biol.* **1997**, 4, 10-19

10. Chan, H. S.; Dill, K. A. Protein Folding in the Landscape Perspective: Chevron Plots and Non-Arrhenius Kinetics. *Proteins* **1998**, 30, 2–33

11. Karplus, M. The Levinthal Paradox: Yesterday and Today. *Fold Des.* **1997**, 2, S69-S75

12. Sali, A.; Shakhnovich, E.; Karplus, M. How does a protein fold? *Nature (London)* **1994**, 369, 248-251

13. Karplus, M; Weaver, D. L. Protein-Folding Dynamics. *Nature* **1976**, 260, 404-406

14. Honeycutt, J. D.; Thirumalai, D., Metastability of the Folded States of Globular Proteins. *Proc. Natl. Acad. Sci. U. S .A.* **1990**, 87, 3526-3529

15. Wolynes P. G.; Onuchic, J. N.; Thirumalai, D. Navigating the Folding Routes. *Science* **1995**, 267, 1619–1620

16. Wales, D. J. The Energy Landscape as a Unifying Theme in Molecular Science. *Phil. Trans. R. Soc. A* **2005**, 363, 357–377

17. Wales D. J.; Bogdan, T. V., Potential Energy and Free Energy Landscapes. *J. Phys. Chem. B* **2006**, 110, 20765-20776

18. Carr, J. M.; Wales, D. J.; Global Optimization and Folding Pathways of Selected -Helical Proteins. *J. Chem. Phys.* **2005**, 123, 23490

19. Baldwin, R. L. The Nature of Protein Folding Pathways: The Classical Versus the New View. *J. Biomol. NMR* **1995**, 5, 103-109

20. Baldwin, R. L. Competing Unfolding Pathways. *Nat. Struct. Biol.* **1997**, 4, 965-966

21. Fersht, A. R.; Itzhaki, L.S.; elMasry, N.F.; Matthews, J. M.; Otzen, D. E. Single versus Parallel Pathways of Protein Folding and Fractional Formation of Structure in the Transition State. *Proc. Natl. Acad. Sci.* **1994**, 91, 10426–10429

22. Fersht, A. R. Optimization of Rates of Protein Folding: The Nucleation-Condensation Mechanism and Its Implications. *Proc. Natl. Acad. Sci. U S A.* **1995** 92, 10869–10873

23. Oliveberg M.; Tan, Y. J.; Fersht, A. R. Negative Activation Enthalpies in the Kinetics of Protein Folding. *Proc. Natl. Acad, Sci, U S A.* **1995**, 92, 8926–8929

24. Pande, V. S. Grosberg, A. Yu.; Tanaka, T.; Rokhsar, D. S. Pathways for Protein Folding: Is a New View Needed? *Curr. Opin. Struct. Biol.* **1998**, 8, 68–79

25. Lindorff-Larsen, K.; Piana, S.; Dror, R. O.; Shaw, D. E. How Fast-Folding Proteins Fold. *Science* **2011**, 334, 517-520

26. Bennion, B. J.; Daggett, V. Counteraction of Urea-Induced Protein Denaturation by Trimethylamine N-Oxide: A Chemical Chaperone at Atomic Resolution. *Proc. Natl. Acad. Sci. U S A.* **2003**, 100, 5142–5147

27. Camilloni, C.; Rocco, A. G.; Eberini, I.; Gianazza, E.; Broglia, R. A.; Tiana, G. Urea and Guanidinium Chloride Denature Protein L In Different Ways In Molecular Dynamics Simulations. *Biophys J.* **2008**, 94, 4654-61

28. Rocco, A.G.; Mollica, L.; Ricchiuto, P.; Baptista, A. M.; Gianazza, E.; Eberini, I. Characterization of the Protein Unfolding Processes Induced by Urea and Temperature. *Biophys J.* **2008**, 94, 2241-51

29. Maity, H.; Maity, M.; Englander, S. W.; How Cytochrome C Folds, and Why: Submolecular Foldon Units and Their Stepwise Sequential Stabilization. *J. Mol. Biol.* **2004**, 343, 223–233

30. Krishna, M. M. G.; Maity, H.; Rumbley, J. N.; Lin, Y.; Englander, S. W.; Order Steps in the Cytochrome C Folding Pathway: Evidence tor a Sequential Stabilization Mechanism. *J. Mol. Biol.* **2006**, 359, 1410–1419

31. Krishna, M. M. G; Englander, S. W. A Unified Mechanism for Protein Folding: Predetermined Pathways with Optional Errors *Protein Science* **2007**, 16, 449–464

32. Kim, S.; Chung, J. K.; Kwak, K.; Bren, K. L.; Bagchi, B.; Fayer, M. D. Native and Unfolded Cytochrome C – Comparison of Dynamics using 2D-IR Vibrational Echo Spectroscopy. *J. Phys. Chem. B.* **2008**, 112, 10054-10063

33. Frank, H. S.; Franks, F. Structural Approach to the Solvent Power of Water for Hydrocarbons; Urea as a Structure Breaker. *J. Chem. Phys.* **1968**, 48, 4746–4757

34. Tobi, D.; Elber, R.; Thirumalai, D. The Dominant Interaction between Peptide and Urea Is Electrostatic in Nature: A Molecular Dynamics Simulation Study. *Biopolymers* **2003**, 68, 359–369

35. O'Brien, E. P.; Dima, R. I.; Brooks, B; Thirumalai, D. Interactions between Hydrophobic and Ionic Solutes in Aqueous Guanidinium Chloride and Urea Solutions: Lessons for Protein Denaturation Mechanism. *J. Am. Chem. Soc*. **2007**,129,7346–7353

36. Das A, Mukhopadhyay C. Urea-Mediated Protein Denaturation: A Consensus View. *J. Phys. Chem. B*. **2009,** 113, 12816–12824

37. Zhou,R.; Li, J.; Hua, L.; Yang, Z.; Berne, B. J.; Comment on "Urea-Mediated Protein Denaturation: A Consensus View". *J Phys Chem B*. **2011**, 115(5),1323–1328

38. Hua, L.; Zhou, R.; Thirumalai, D.; Berne, B. J. Urea Denaturation by Stronger Dispersion Interactions with Proteins than Water Implies a 2-Stage Unfolding. *Proc. Natl. Acad. Sci. U.S.A.* **2008**, 105, 16928–16933

39. Caflisch, A.; Karplus, M. Structural Details of Urea Binding to Barnase: A Molecular Dynamics Analysis. *Structure* **1999**, 7, 477–488

40. Roy, S.; Jana, B.; Bagchi, B. Dimethyl Sulfoxide Induced Structural Transformations and Non-Monotonic Concentration Dependence of Conformational Fluctuation around Active Site of Lysozyme. *J. Chem. Phys.* **2012**, 136, 115103/1-10

41. Jackson, M.; Mantsch, H. H. Beware of proteins in DMSO. *Biochim. Biophys. Acta.* **1991**, 1078, 231-5

42. Huang, P.; Dong, A.; Caughey, W. S. Effects of Dimethyl Sulfoxide, Glycerol, and Ethylene Glycol on Secondary Structures of Cytochrome C and Lysozyme as Observed by Infrared Spectroscopy. *J. Pharm. Sci.* **1995**, 84, 387-392

43. Roy, S.; Bagchi, B. Chemical Unfolding of Chicken Villin Headpiece in Aqueous Dimethyl Sulfoxide Solution: Cosolvent Concentration Dependence, Pathway, and Microscopic Mechanism. *J. Phys. Chem B* **2013**, 117, 4488-4502

44. Tretyakova, T.; Shushanyan, M.; Partskhaladze, T.; Makharadze, M.; van Eldik, R.; Khoshtariya, D. E. Simplicity within the Complexity: Bilateral Impact of DMSO on the

Functional and Unfolding Patterns of α-Chymotrypsin. *Biophys. Chem.* **2013**, 175-176, 17-27

45. Batista, A. N.; Batista, J. M. Jr; Bolzani, V. S.; Furlan, M.; Blanch, E. W. Selective DMSO-Induced Conformational Changes in Proteins from Raman Optical Activity. *Phys. Chem. Chem. Phys.* **2013**, 15, 20147-20152

46. Piana, S.; Lindorff-Larsen, K.; Shaw, D. E. Atomic-Level Description of Ubiquitin Folding. *Proc. Natl. Acad. Sci. U S A*. **2013**, 110, 5915-5920

47. Larios, E.; Yang, W. Y.; Schulten K.; Gruebele, M. A Similarity Measure for Partially Folded Proteins: Application to Unfolded and Native-Like Conformation Fluctuations. *Chem. Phys*. **2004**, 307, 217–225

48. Yang, W. Y; Pitera, J. W; Swope, W. C; Gruebele, M. Heterogeneous Folding of the trpzip Hairpin: Full Atom Simulation and Experiment. *J. Mol. Biol.* **2004**, 336, 241-251

49. Larios, E.; Li, J. S.; Schulten, K.; Kihara, H.; Gruebele, M. Multiple Probes Reveal a Native-Like Intermediate During Low-Temperature Refolding of Ubiquitin *J. Mol. Biol.* **2004**, 340, 115-125

50. Gronenborn, A.M.; Filpula, D.R.; Essig, N.Z.; Achari, A.; Whitlow, M.; Wingfield, P.T.; Clore, G.M. A Novel, Highly Stable Fold of the Immunoglobulin Binding Domain of Streptococcal Protein G. *Science* **1991**, 253, 657-661

51. McPhalen, C.A.; James, M. N. Crystal and Molecular Structure of the Serine Proteinase Inhibitor CI-2 from Barley Seeds. *Biochemistry* **1987**, 26, 261-269

52. Wikstrom, M.; Drakenberg, T.; Forsen, S.; Sjobring, U.; Bjorck, L. Three-Dimensional Solution Structure of an Immunoglobulin Light Chain-Binding Domain of Protein L. Comparison with the Igg-Binding Domains Of Protein G. *Biochemistry* **1994**, 33, 14011-14017.

53. Vijay-Kumar, S.; Bugg, C.E.; Cook, W.J. Structure of Ubiquitin Refined at 1.8 Å Resolution. *J.Mol.Biol.* **1987**, 194, 531-544

54. Oostenbrink, C.; Villa, A.; Mark, A. E.; van Gunsteren, W. F. A Biomolecular Force Field Based on the Free Enthalpy of Hydration and Solvation: The GROMOS Force-Field Parameter Sets 53A5 and 53A6. *Comput. Chem*. **2004**, 25, 1656-1676

55. Liu, H. Y.; Müller–Plathe, F.; van Gunsteren, W. F. A Force Field for Liquid Dimethyl Sulfoxide and Physical Properties of Liquid Dimethyl Sulfoxide Calculated Using Molecular Dynamics Simulation. *J. Am. Chem. Soc*. **1995**, 117, 4363-4366

56. Geerke, D. P.; Oostenbrink, C.; van der Vegt, N. F. A.; van Gunsteren, W. F. An Effective Force Field for Molecular Dynamics Simulations of Dimethyl Sulfoxide and Dimethyl Sulfoxide−Water Mixtures. *J. Phys. Chem. B* **2004**, 108, 1436-1445

57. Smith, L. J.; Berendsen, H. J. C.; van Gunsteren, W. F. Computer Simulation of Urea-Water Mixtures: A Test of Force Field Parameters for Use in Biomolecular Simulation. *J Phys Chem B* **2004**, 108,1065-1071

58. Berendsen, H. J. C.; Grigera, J. R.; Straatsma, T. P. The Missing Term in Effective Pair Potentials. *J. Phys. Chem*. **1987**, 89, 6269-6271

59. Hoover, W. G. Canonical Dynamics: Equilibrium Phase-Space Distributions. *Phys. Rev. A* **1985**, 31, 1695-1697

60. Nose, S. A Unified Formulation of the Constant Temperature Molecular Dynamics Methods. *J. Chem. Phys*. **1984**, 81, 511-519

61. Parinello, M.; Rahman, A. Polymorphic Transitions in Single Crystals: A New Molecular Dynamics Method. *J. Appl. Phys*. **1981**, 52, 7182-7190

62. Frenkel, D.; Smit, B. Understanding Molecular Simulation: From Algorithms to Applications (2nd ed., Academic Press: San Diego, CA) **2002**

63. Roy, S.; Banerjee, S.; Biyani, N.; Jana, B.; Bagchi, B. Theoretical and computational analysis of static and dynamic anomalies in Water−DMSO binary mixture at low DMSO concentrations. *J. Phys. Chem. B* **2010**, 115, 685-692

64. Banerjee, S.; Roy, S.; Bagchi, B. Enhanced pair hydrophobicity in water – DMSO binary mixture at low DMSO concentration *J. Phys. Chem. B* **2010**, 114, 12875-12882

65. Roy, S.; Bagchi, B. Solvation dynamics of tryptophan in water-dimethyl sulfoxide binary mixture: In search of molecular origin of composition dependent multiple anomalies *J. Chem. Phys*. **2013**, 139, 034308/1-10